\documentclass[amsmath, amssymb, twocolumn, nofootinbib, superscriptaddress, showkeys, aps, prb, 10pt]{revtex4-2}
\usepackage{header}

\begin{document}
\title{High-Energy Interlayer Exciton Ensembles in \texorpdfstring{MoSe$_2$}{MoSe2}/\texorpdfstring{WSe$_2$}{WSe2} Heterostructures by Laguerre-Gaussian Excitation}
\def\wsi{Walter Schottky Institute and Physics Department, TU Munich, Am Coulombwall 4a, 85748 Garching, Germany}
\def\mcqst{Munich Center for Quantum Science and Technology (MCQST), Schellingstr. 4, 80799 Munich, Germany}
\def\berlin{Institute for Theoretical Physics, Nonlinear Optics and Quantum Electronics, Technical University of Berlin, 10623 Berlin, Germany}
\def\muenster{Institute of Physics, University of Münster, Wilhelm-Klemm-Str. 10, 48149 Münster, Germany}
\def\mana{Research Center for Materials Nanoarchitectonics, National Institute for Materials Science, 1-1 Namiki, Tsukuba 305-0044, Japan}
\def\nims{Research Center for Electronic and Optical Materials, National Institute for Materials Science, 1-1 Namiki, Tsukuba 305-0044, Japan}

\author{Mirco Troue}

\author{Johannes Figueiredo}
\author{Gabriel Mittermair}
\author{Jonas Kiemle}
\author{Sebastian Loy}\affiliation{\wsi}\affiliation{\mcqst}
\author{Hendrik Lambers}\affiliation{\muenster}


\author{Takashi Taniguchi}\affiliation{\mana}
\author{Kenji Watanabe}\affiliation{\nims}

\author{Ursula Wurstbauer}\affiliation{\muenster}
\author{Alexander W. Holleitner}\email[Electronic address: ]{holleitner@wsi.tum.de}\affiliation{\wsi}\affiliation{\mcqst}
\date{\today}

\begin{abstract}
We reveal the higher energetic luminescence part of interlayer exciton ensembles in MoSe$_2$/WSe$_2$ heterostructures upon excitation by an optical Laguerre-Gaussian mode. The excitation is achieved with the help of a spatial light modulator giving rise to a ring-shaped distribution of interlayer excitons. A hyperspectral analysis of the exciton photoluminescence suggests that the excitation scheme allows the accumulation of high-energetic excitons in the rings' center. We discuss the mechanisms leading to such a distribution, including exciton-exciton interaction, phase-space filling, and an incomplete thermalization.
\end{abstract}


\maketitle

\section{Introduction}
In the rapidly growing family of two-dimensional materials, spatially indirect excitons with strong light-matter coupling play an important role in the current research efforts to develop next-generation optoelectronic quantum devices \cite{Brotons-Gisbert2024, Rivera2015, miller_long-lived_2017,wilson_determination_2017,hanbicki_double_2018, Unuchek2018, Jauregui2019, Qian2024, Figueiredo2025}. The Coulomb-bound interlayer excitons, formed e.g. across the interface of a \ce{MoSe2}/\ce{WSe2} van-der-Waals heterostack, exhibit unique properties such as a permanent out-of-plane electric dipole moment and a large exciton binding energy of several hundreds of \si{\milli\electronvolt} \cite{Brotons-Gisbert2024, Rivera2015}. Long photoluminescence lifetimes on the order of \SI{100}{ns} make them suitable for investigating many-body phenomena in two-dimensional exciton ensembles in low dimensions \cite{Fogler2014}. Recent reports on many-body effects of such interlayer excitons (IXs) suggest an excitonic degeneracy with a macroscopic spatial coherence at sufficiently high exciton densities and low temperatures \cite{Sigl2020, Troue2023, Katzer2023, Cutshall2025, Fogler2014}.

Spatially homogeneous ensembles of IXs are typically formed by photoexciting intralayer excitons in the monolayers of the heterostructures in combination with interlayer hybridization, which leads to an efficient charge transfer into the interlayer configuration \cite{madeo_directly_2020, dong_direct_2021, Schmitt2022}. The resulting IXs undergo ultrafast relaxation and thermalization \cite{Policht2023}, and in turn, a Gaussian laser profile results in a corresponding two-dimensional isotropic ensemble of IXs. Time-resolved transient-absorption measurements on such ensembles reveal pronounced non-thermal spectral signatures whose temporal evolution reflects cooling e.g. via LO phonons \cite{Jiang2021, Yoon2022, Policht2023}. Using a two-dimensional coherent spectroscopy, the cooling dynamics of both intra- and interlayer excitons in heterostructures can be resolved on ultrafast timescales \cite{Purz2021, Barre2024}. However, to date, the separation, detection, and analysis of non-thermalized, high-energy exciton ensembles in continuously pumped \ce{MoSe2}/\ce{WSe2} heterobilayers has not been achieved. Long-lived, but non-thermalized IXs are supposed to play an important role in the formation and behavior of interlayer exciton ensembles extending beyond the excitation spot, which are expected to host correlation phenomena \cite{Schinner2011, Troue2023, Cutshall2025}.

Here, we introduce a Laguerre-Gaussian (LG) excitation scheme as created with the help of a spatial light modulator to generate IXs in \ce{MoSe2}/\ce{WSe2} heterostructures encapsulated in hexagonal boron nitride (\ce{h-BN}). Such optical LG excitations are widely used in exciton-polariton experiments where lasing \cite{Paik2019, Qian2024} and polariton condensation \cite{Kasprzak2006} have been reported. We demonstrate that this excitation scheme allows to form ring-shaped ensembles of IXs with a tunable diameter on the order of a few micrometers as well as with differing properties at the perimeter and the center of the ring. In the center of the rings, we find an increasing exciton photoluminescence (PL) with respect to the position on the ring as a function of excitation power. Moreover, we observe that the central exciton luminescence exhibits a broadening, particularly on the high-energy tail. We interpret both the increased center PL intensity and the spectral broadening in terms of non-thermalized IX propagating from the perimeter of the ring into the center. The demonstrated LG excitation scheme opens the pathway to form IX ensembles in unprecedented shapes and to spatially separate non-thermalized IXs from their excitation.

\begin{figure}[ht]
    \centering
    \includegraphics[width=\columnwidth]{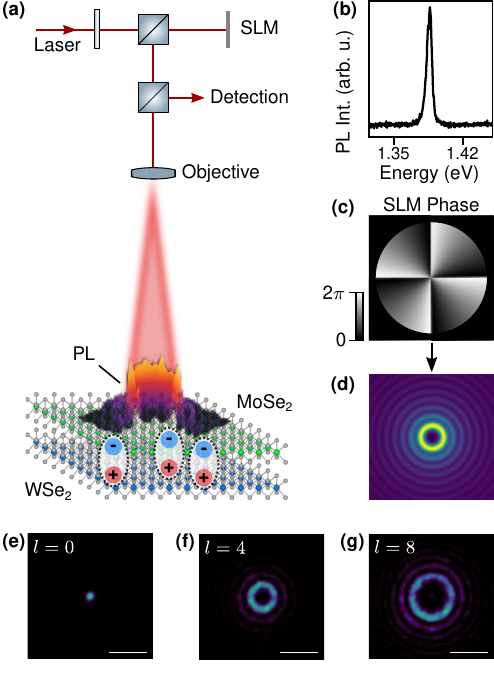}
    \caption{Laguerre-Gaussian (LG) excitation scheme of interlayer excitons (IXs) in a \ce{MoSe2}/\ce{WSe2} heterostructure. (a) A spatial light modulator (SLM) reflects the phase-modulated laser light back into the optical circuitry. With the help of a beam splitter and an objective, the light is then focused onto the sample. (b) IX PL at an excitation power of \SI{50}{\nano\watt} ($E_{\mathrm{Laser}} = \SI{1.94}{\electronvolt}$, $T_{\mathrm{Bath}} = \SI{1.7}{\kelvin}$). (c) Spatial phase pattern of the SLM. Black overlay sketches the circular aperture of the optical circuitry. (d) Computed intensity pattern in the far-field corresponding to the phase pattern of (b). (e)-(g): Excitation laser profiles, as reflected from the samples' substrate and measured with a camera in the detection path for topological orders (e): $l = 0 $, (f): $4$, and (g): $8$. Scale bars, \SI{10}{\micro\meter}.}
    \label{fig:1}
\end{figure}

\section{Results}
We investigate \ce{MoSe2}/\ce{WSe2} heterostructures encapsulated in \ce{h-BN}, where the optically generated IXs comprise a photo-generated electron in the \ce{MoSe2} layer and a corresponding hole in the \ce{WSe2} layer [cf. \cref[(a)]{fig:1}]. The heterostacks are placed on top of \ce{Si}/\ce{SiO2} substrates and characterized with the help of a cryogenic PL setup (Methods). The PL of the heterostacks, as excited at $E_{\mathrm{Laser}} = \SI{1.94}{\electronvolt}$ with \SI{50}{\nano\watt}, features a sharp emission at an energy of around $E_{\mathrm{IX}} =\SI{1.4}{\electronvolt}$ with a full width at half maximum of \SI{5.5}{\milli\electronvolt} at \SI{1.7}{\kelvin} [cf. \cref[(b)]{fig:1}]. Consistent with the literature \cite{Brotons-Gisbert2021, Rivera2015, Figueiredo2025, Fogler2014, Sigl2020, Troue2023, Cutshall2025, Ross2017, Nayak2017}, we interpret this PL maximum to stem from the mentioned IXs.
\Cref[(c)]{fig:1} depicts a phase pattern of the utilized spatial light modulator (SLM) with four phase quadrants with phase shifts ranging from 0 to 2$\pi$, which allows creating a Laguerre-Gaussian (LG) mode with $l = 4$ in the far field [cf. \cref[(d)]{fig:1} and Methods]. \Crefs[(e)-(g)]{fig:1} show the SLM-modified laser profiles as reflected from the silicon substrate of sample 1 for topological orders $l=0$, $4$, and $8$. For $l=0$, the reflection profile is a simple Gaussian with a width of \SI{785(2)}{\nano\meter}, and it resembles the point spread function of the optical circuit. For $l=4$, we observe a ring-shaped profile with a negligible intensity in the center, while the diameter of the main reflection ring equals $d = \SI{4.6}{\micro\meter}$ [cf. Supporting Fig. S1]. The size of the main excitation ring for this order suits the lateral extension of the investigated heterostructures on the order of a few to ten micrometers. In turn, we choose this ring configuration ($l = 4$ and $ d = \SI{4.6}{\micro\meter}$) for the rest of the experiments.

\Crefs[(a) and (b)]{fig:2} display the emitted PL pattern upon excitation with the aforementioned ring-shaped excitation profile ($l=4$). For this experiment, we image the PL of the heterostructures as a function of the total excitation power with the help of a CMOS camera. Moreover, we utilize a combination of optical filters, such that the CMOS camera only detects the PL of the IXs [cf. Fig. 1(b) and Methods]. \crefs[(a) and (b)]{fig:2} showcase the possibility of exciting IX ensembles in circular patterns for two laser powers. We clearly observe a circular PL profile visible for all excitation powers. As a reference, the bottom panel of \cref[(c)]{fig:2} shows a line cut through the reflected laser illumination profile. The two main maxima, highlighted by a gray background in \cref[(c)]{fig:2}, mark the spatial region with high excitation power density from the LG mode. Moreover, the illumination profile features a rather negligible intensity in the center ($\blacktriangledown$). The top panel of \cref[(c)]{fig:2} displays exemplary line cuts of the PL intensities along the dashed line in \cref[(a)]{fig:2} for various excitation powers. As expected, the exciton PL is strongest at the position of the main illumination maxima [one maximum is exemplarily highlighted by an open circle ($\circ$)]. In a next step, we compare the PL at the center of the excitation ($\triangledown$) to the one at a distance ($\square$), which is equally spaced away from the ring ($\circ$). Interestingly, the center PL ($\triangledown$) exceeds the one outside of the ring ($\square$), albeit the excitation at both positions is comparable [cf. laser intensity at positions marked by $\blacktriangledown$ and $\blacksquare$ in the bottom panel of \cref[(c)]{fig:2}]. Tentatively, we explain the increased center PL by an increased IX density because of diffusing and drifting IXs towards the ring center, while the overall decay time remains unchanged, as will be discussed in detail below.

\begin{figure}[ht]
    \centering
    \includegraphics[width=\columnwidth]{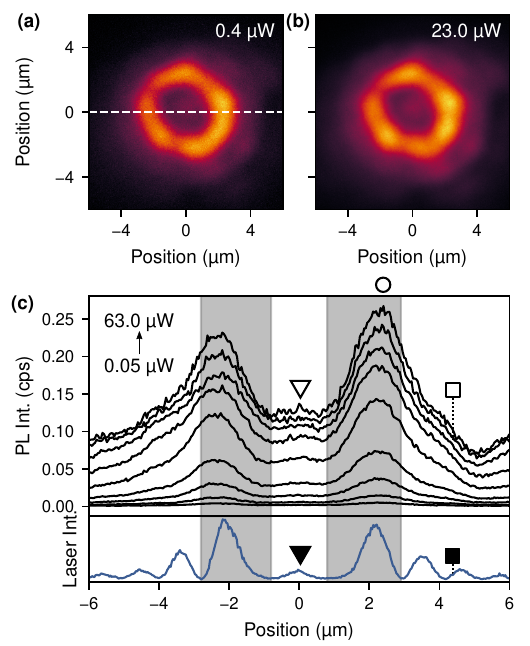}
    \caption{Ring-shaped pattern of IX PL from sample 1. (a) and (b): Camera images of the circular PL pattern for two different excitation powers ($E_{\mathrm{Laser}} = \SI{1.94}{\electronvolt}$, $T_{\mathrm{Bath}} = \SI{1.7}{\kelvin}$). (c) PL profiles along the dashed line in (a) with the excitation power ranging from \SI{50}{\nano\watt} to \SI{63}{\micro\watt}. Bottom panel depicts a corresponding line cut through the reflected laser illumination profile. Triangles mark the center of the excitation ($\blacktriangledown$) and the PL ($\triangledown$). Circle ($\circ$) highlights one of the PL maxima on the ring (gray shaded). Square ($\square$) marks a position outside of the ring that is equally spaced away from the excitation maximum as the center.}
    \label{fig:2}
\end{figure}

\begin{figure*}
    \centering
    \includegraphics[width=2\columnwidth]{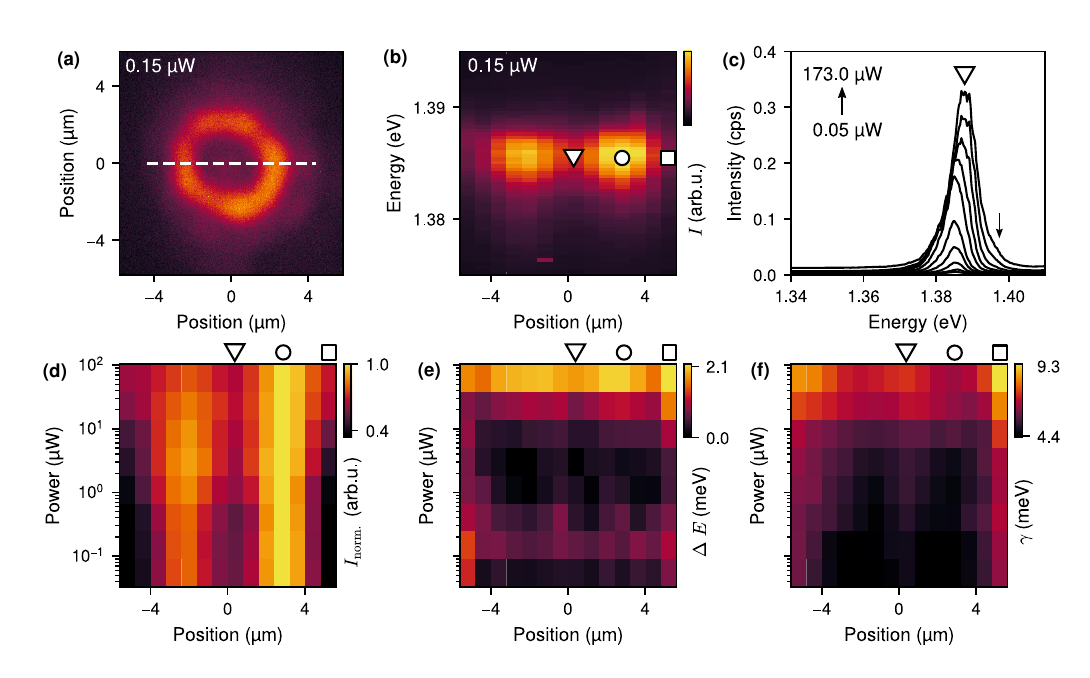}
    \caption{Hyperspectral characterization of the IX PL on sample 1. (a) PL image in the sample plane recorded by a camera (similar to Fig. 2). (b) Spatio-spectral scan along the dashed line in (a) with the help of a glass fiber in the detection path (cf. Fig. 1(a) and Supporting Information). (c) Corresponding IX PL spectra for varying excitation powers collected in the center of the PL pattern as indicated by the triangle ($\triangledown$) in (b). A small arrow indicates the onset of an additional high-energy peak arising at high excitation powers. (d)-(f): Intensity $I_{\mathrm{norm.}}$ normalized to the intensity on the ring ($\circ$), energy shift $\Delta E$, and linewidth $\gamma$ of the IX PL extracted from Lorentzian fits of the acquired spectra for varying position and excitation power [cf. positions $\triangledown$, $\circ$, and $\square$ in Fig. \sref[(b) and (c)]{fig:2}].}
    \label{fig:3}
\end{figure*}

To further investigate the power-dependent PL characteristics of the IX ensemble, we perform a hyperspectral analysis of the IX rings' PL. To this end, we collect the emitted light in the image plane of the detection path with the help of an optical fiber and analyze it by utilizing a spectrometer. \cref[(a)]{fig:3} shows the original PL image as measured by the CMOS camera, while \cref[(b)]{fig:3} provides a corresponding spectrally resolved scan along the dashed line in \cref[(a)]{fig:3} [cf. Supporting Figs. S3 and S4]. There are two PL maxima [one marked with $\circ$ in \cref[(b)]{fig:3}]. Both correspond to the positions of the ring excitation, while the PL intensity at the center ($\triangledown$) and outside of the ring ($\square$) is reduced. \Cref[(c)]{fig:3} depicts the PL spectra measured at the center ($\triangledown$) for several excitation powers. For all powers, we observe a sharp single PL peak with a moderate blue shift towards higher energies with increasing laser power. Only at the highest powers, a new peak emerges in the high-energy shoulder of the PL [arrow in \cref[(c)]{fig:3}], as it is consistent with previous reports on IXs in such heterostructures \cite{Sigl2020, Sigl2022, Figueiredo2025}. For the rest of the manuscript, we restrict ourselves to the excitation regime where only one PL maximum appears in the spectra (powers smaller than \SI{63}{\micro\watt} for $l = 4$). We fit each spectrum with the help of a Lorentzian function and extract the amplitude, the energy shift with respect to the lowest peak position, and the linewidth of the PL as a function of the laser power . \Cref[(d)]{fig:3} depicts the fitted amplitude normalized to the PL at the right ring maximum ($\circ$) as a function of position and the experimental excitation power. For the center position ($\triangledown$), we observe an increasing PL, which exceeds the PL at positions outside of the ring ($\square$). The increase in the center intensity is accompanied by a blue shift of up to \SI{2.1}{\milli\electronvolt} as shown in \cref[(e)]{fig:3}. We note that, particularly for low excitation powers, the energy shift is independent of the position, which suggests an interaction and/or exchange of IX across the interior of the ring. For the highest excitation powers, the blue shift is slightly increased at the main excitation maxima ($\circ$). Stronger exciton-exciton interactions likely cause this effect because of a higher exciton density on the ring~\cite{steinhoff_exciton-exciton_2024}. The difference in blue shift amounts to a value of \SI{0.36}{\milli\electronvolt} between the ring maximum and center at the highest excitation powers. \Cref[(f)]{fig:3} displays the extracted linewidth of the PL. For all laser powers, the scans feature an increased linewidth in the center ($\triangledown$) compared to the positions where the excitation occurs ($\circ$). We note that the positions outside of the ring (e.g. $\square$) also feature a higher linewidth, as will be discussed below.

\begin{figure}
    \centering
    \includegraphics[width=\columnwidth]{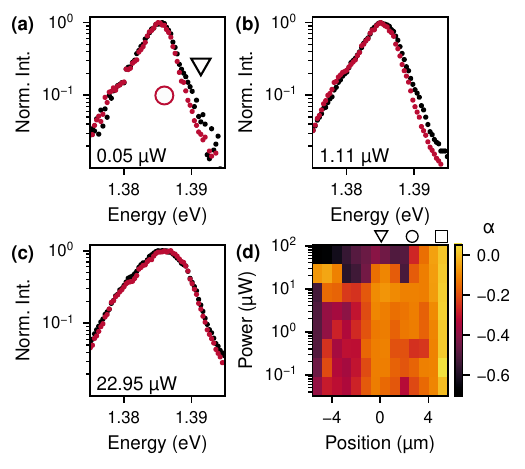}
    \caption{Comparison of an IX PL spectrum at the ring center ($\triangledown$, black) to a spectrum at the ring maximum ($\circ$, red) for sample 1. (a) Logarithmic presentation of the normalized intensities for an excitation power of \SI{50}{\nano\watt}. (b) and (c): Similar presentations for \SI{1.11}{\micro\watt} and \SI{22.95}{\micro\watt}. (d) Skewness shape parameter $\alpha$ for the hyperspectral data of \cref{fig:3}. See text for details.}
    \label{fig:4}
\end{figure}

To explore the increased linewidth of the PL in the center compared to the one at the ring in greater detail, \crefs[(a)-(c)]{fig:4} show individual spectra of the hyperspectral data from \cref{fig:3} in a logarithmic plot. Indeed, at an excitation power of \SI{50}{\nano\watt} and \SI{1.11}{\micro\watt} [\crefs[(a) and (b)]{fig:4}], the PL features a more pronounced high-energy tail at the center position ($\triangledown$) compared to the position on the ring ($\circ$). A further increase to \SI{22.95}{\micro\watt} reduces this effect [\cref[(c)]{fig:4}]; most likely, because the spectrum broadens with higher powers, since more and more photo-excited IXs, intralayer excitons, and charge carriers are introduced to the system \cite{Moody2015, Katzer2023}. To further investigate the high-energy tail of the PL at the ring center, we utilize a skewed Lorentzian fit as defined in \cref{eq:skewed} and apply it to all spectra (cf. Supporting Fig. S5).

\begin{equation}
    \dfrac{A}{\pi} \dfrac{\gamma / 2}{\left(E-E_{\mathrm{IX}}\right)^2 + \left(\gamma/2\right)^2} \left(1+\mathrm{erf}\left( \alpha \dfrac{\left(E-E_{\mathrm{IX}}\right)}{\gamma/2}\right)\right) + c,
    \label{eq:skewed}
\end{equation}

where $A$ describes the amplitude of the Lorentzian, $\gamma$ the FWHM, and $c$ the background. Here, the shape parameter $\alpha$ inside the error function skews the weight of the Lorentzian to either lower ($\alpha < 0$) or higher energies ($\alpha > 0$), while a setting with $\alpha = 0$ yields a symmetric Lorentzian. \Cref[(d)]{fig:4} depicts the extracted values of $\alpha$ as a function of position and excitation power. Generally, we observe that $\alpha$ is negative for all positions. This indicates a sharper drop of the PL towards the high-energy side compared to the low-energy side, consistent with previous reports reporting a low-energy tail arising from various contributions such as localized exciton states and phonons \cite{Figueiredo2025}. Still, $\alpha$ is less negative at the center ($\triangledown$) compared to the position on the ring ($\circ$). This increase of $\alpha$ corresponds to an increased high-energy tail of the PL, as discussed already in connection with single spectra presented in \crefs[(a) and (b)]{fig:4}. Equally, outside of the ring ($\square$), $\alpha$ is less negative than at the center position. At excitation powers above \SI{10}{\micro\watt}, the relative high-energy broadening of the PL seems to be reduced again, as it is consistent with generally broadened spectra as in \cref[c]{fig:4}.

We note that the SLM setup allows switching to a Gaussian excitation mode with $l = 0$ without any change of the optical components [cf. \crefs[(a), (e) - (f)]{fig:1}]. In the Supporting Fig. S6, we present corresponding individual and hyperspectral data for a Gaussian excitation on the very same position as the ring experiments for $l = 4$ are performed. For a Gaussian excitation, a single intensity maximum is recovered at the center position with a consistently increased spatial expansion of the IX ensemble at high excitation powers. Again, the blue shift increases position-independently with increasing power for low excitation powers. Moreover, the PL emission away from the excitation spot exhibits an increased high-energy tail, as it is consistent with the PL at the center position and outside of the ring.

\begin{figure}
    \centering
    \includegraphics[width=\columnwidth]{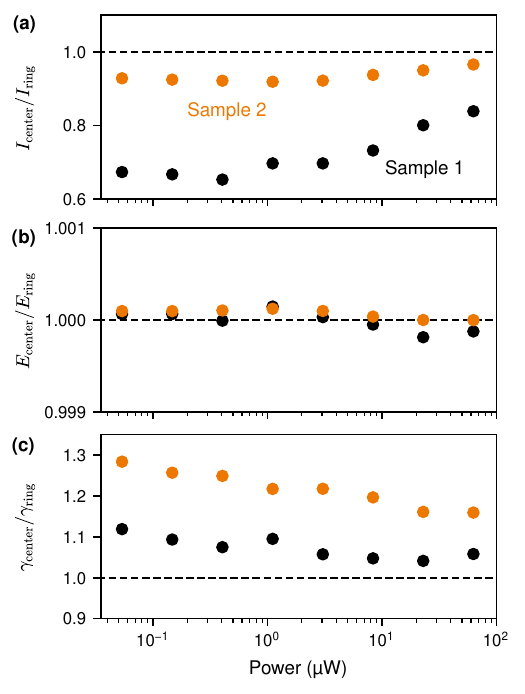}
    \caption{Comparison of intensity, emission energy, and linewidth for the center position to values at the ring maximum for sample 1 (black) and 2 (orange). (a) Ratio of IX PL intensity $I_{\mathrm{center}}/I_{\mathrm{ring}}$ vs. excitation power. (b) Ratio of PL emission energy $E_{\mathrm{center}}/E_{\mathrm{ring}}$ vs. excitation power. (c) Ratio of PL linewidth $\gamma_{\mathrm{center}}/\gamma_{\mathrm{ring}}$ vs. excitation power. }
    \label{fig:5}
\end{figure}
\Cref{fig:5} compares (a) the intensity, (b) the emission energy, and (c) the linewidth of the PL as derived at the ring center to the values at the ring excitation for two samples (both with $l = 4$). Measurements of sample 1 are presented in \cref{fig:1} - \cref{fig:4}, while the corresponding data of sample 2 are presented in the Supporting Information. Strikingly, both samples show a relative increase in intensity at the center for increasing powers [cf. \cref[(a)]{fig:5}]. We note again that all data are in the regime with only one PL maximum in the spectra, such as in \cref[(b)]{fig:1} and \cref[(c)]{fig:3}.
The higher relative center intensity for sample 2 compared to the intensity for sample 1 can be attributed to a spatially more inhomogeneous PL pattern for sample 2 (cf. Supporting Information). For both samples, however, the relative emission energy seems to be constant within the investigated range of powers [cf. \cref[(b)]{fig:5}], indicating a rather position-independent blue shift across the ring pattern [cf. \cref[(e)]{fig:3}]. The relative linewidth, as shown in \cref[(c)]{fig:5}, remains persistently above a value of 1, indicating an effectively broader linewidth in the center of the trap compared to the position on the excitation ring. For both samples, the broadest relative linewidth is observed in the low-power regime, and the relative linewidth drops with increasing powers. Again, this trend is consistent with the spectra as shown in \cref{fig:4}. Last but not least, we note that although the relative intensity at the center of the ring is largest at high exciton powers [\cref[(a)]{fig:5}], the difference in linewidth is best detected at low excitation powers [\cref[(c)]{fig:5}].

\section{Discussion}
With the possibility to directly compare the LG excitation ($l \neq 0 $) to the Gaussian excitation mode ($l = 0$), we can distinguish position-dependent effects that might arise from impurities, disorder, or strain within the heterostructure from effects that are induced by the ring-shaped excitation pattern, such as IX density and effective temperature gradients. A key observation is that for the investigated range of distances, the linewidth of the PL increases with a larger spatial distance from the excitation position for both, the ring [cf. \cref[(f)]{fig:3}] and for the Gaussian excitation geometry [cf. Supporting Fig. S6]. In both cases, the broadening occurs on the high-energy tail of the PL [cf. low excitation regime in \cref{fig:4} and Supporting Fig. S6], indicating a larger contribution of high-energy states to the overall PL at farther propagation distances. Intriguingly, for the discussed ring ($l = 4$) and the Gaussian mode ($l = 0$), the positions of excitation and detection swap in the lateral direction: for the ring, the high-energy broadening of interest occurs at the center with the excitation being at the ring perimeter, while for the Gaussian mode, the excitation is located at the center with an high-energy broadening being detectable radially away from it. In turn, we can exclude the position-dependent effects based on the inhomogeneity of the heterostructure to explain the observed broadening at the ring center. A second observation is that the blue shift of the PL energy varies only negligibly across the IX ring area for the investigated range of powers [cf. \crefs[(e)]{fig:3} and \sref[(b)]{fig:5}]. In turn, neither spatially varying exciton-exciton interactions nor a local phase-space filling of lower energy states alone can be responsible for the broadening of the emission to higher energies at the ring center. Instead, we explain the observed effects by an incomplete thermalization of the IXs propagating to the center of the ring.
Generally, IXs are reported to laterally propagate across several hundreds of nanometers up to even micrometers in certain geometrical settings, materials, and temperatures \cite{vogele_density_2009, Rossi2024, Wietek2024, Jauregui2019, fowler-gerace_transport_2024, unuchek_room-temperature_2018}. For the specific ring-shaped excitation, one can see this IX propagation in \cref[(c)]{fig:2}, where the PL intensity profile is spatially broader than the excitation profile. The observed spatial broadening is on the order of \SI{800}{\nano\meter}, which is consistent with previous reports on the diffusion of IXs \cite{Rossi2024, Wietek2024, Jauregui2019, fowler-gerace_transport_2024, unuchek_room-temperature_2018}. 

The Supporting Fig. S7 shows the experimentally determined decay time of the excitons, which does not vary across the ring geometry within the given signal-to-noise ratio. This indicates that the observed changes in PL intensity do not originate from a varying exciton lifetime, but rather from a change in IX density. Moreover, the discussed effects are observed in a regime where the PL decay is mono-exponential. We therefore conclude that the emission is dominated by one type of IXs, as consistent with the PL spectra [low excitation regime in Fig.~\ref{fig:3}(c)]. 

Combining all of the above arguments, the specificity of our experiments is the ring-shaped excitation of the IX ensembles (for $l \neq 0$). In our understanding, it explains the increased PL intensity in the center of the rings compared to positions outside of the ring [e.g. position $\triangledown$ vs $\square$ in \cref[(c)]{fig:2}] and the relative increase of the center PL wrt. the PL on the ring as a function of excitation power [cf. \cref[(a)]{fig:5}]. In other words, the ring geometry focuses the propagation of non-thermalized IX into the center, where they give rise to an overall increased PL compared to positions outside of the ring [e.g. compare the PL intensity at positions $\triangledown$ and $\square$ in \cref[(c)]{fig:2}]. We note that the overall dynamics might be further driven and influenced by the repulsive permanent out-of-plane electric dipole moment of the IXs \cite{kiemle_control_2020, Erkensten2022}, and the spatial extent of the IX wavefunctions as well as the relaxation pathways via intra- and interlayer exciton states \cite{Troue2023, Figueiredo2025}.
Moreover, the observed broadening effect on the high-energy side of the PL emission is generally limited by the accessible light cone as given by the utilized optical circuitry. At higher laser powers, more phonons are present that facilitate the faster thermalization of excitons to lower states \cite{Brem2018, Policht2023}, and photo-generated charge carriers progressively screen Coulomb interactions, such that the exciton binding energy is reduced \cite{Xiao2024}. Both the presence of phonons and a possible faster dissociation of excitons explain that the broadening effect levels off at a higher excitation power [cf. \crefs[(a)-(c)]{fig:4} and \sref[(c)]{fig:5}].

We finally note that the LG modes with $l \neq 0$ might excite higher-momentum intra- and interlayer exciton states, which can result in a PL comprising also an out-of-plane polarization. Back-focal plane experiments, however, demonstrated that the discussed IXs feature a rather pure in-plane optical transition dipole \cite{Sigl2022}. In turn, a PL detection at oblique angles might allow to investigate the expansion and propagation characteristics of possible higher-momentum IX states in the demonstrated ring-shaped geometries. At the current state, such experiments are beyond the scope of the presented work.

\section{Conclusion}
In summary, we investigate the photoluminescence (PL) of ring-shaped ensembles of interlayer excitons (IXs) in \ce{MoSe2}/\ce{WSe2} heterostacks, which are encapsulated in \ce{h-BN}. We create the ring-shaped excitation patterns in terms of Laguerre-Gaussian modes with the help of a spatial light modulator (SLM). The implementation of the SLM into the excitation path of the optical circuitry directly enables confocal detection of the exciton PL without the influence of the excitation laser. In the center of the rings, we find an increasing exciton PL intensity with respect to the position on the rings as a function of excitation power. Moreover, we observe that the centered exciton luminescence exhibits a spectral broadening, particularly on the high-energy tail. We interpret both the increased center luminescence and the broadening in terms of non-thermalized IX propagating from the perimeter of the rings into the center. Future experiments might combine the demonstrated non-Gaussian excitation profiles with a momentum-resolved and coherent detection to investigate the relaxation and coherence mechanisms in ring-shaped exciton ensembles with separated thermalized and non-thermalized contributions \cite{Sigl2020, Sigl2022, Troue2023}.

\section{Methods}
\subsection{Sample preparation}
The investigated heterostructures consist of micromechanically exfoliated monolayers of \ce{MoSe2} and \ce{WSe2} stacked with a twist angle close to \SI{60}{\degree}. After exfoliation, we use viscoelastic stamping to encapsulate the monolayers in \ce{h-BN} and position them on a \ce{Si}/\ce{SiO2}-substrate. The heterostacks are vacuum-annealed at \SI{400}{\kelvin} for two hours.

\subsection{Generation of the circular excitation mode}
We use a spatial light modulator (SLM) based on a phase-only Liquid Crystal on Silicon (LCoS) to shape a collimated linearly-polarized excitation laser beam into a Laguerre-Gaussian (LG) mode by imprinting a specific phase pattern onto the wavefront [cf. Figs. \ref{fig:1}(a) and (c)]. The phase pattern is designed such that the reflected beam acquires a helical phase structure, characteristic of the LG mode. \Cref[(c)]{fig:1} depicts a phase pattern with four phase quadrants with phase shifts ranging from 0 to $2 \pi$, which allows creating an LG mode with $l=4$ in the far field. The number of phase jumps defines the topological order $l$ of the LG mode. Fourier transforming the phase pattern yields a radial intensity profile with a minimum at the center [\cref[(d)]{fig:1}]. The outer circles with lower intensity arise from the impact of the circular aperture in the optical circuitry [cf. black area in \cref[(b)]{fig:1}]. A mirror located behind the LCoS inside the SLM reflects the phase-modulated light back onto a beam splitter, which guides the light onto the sample. We use an objective to create a diffraction-limited image on the sample that corresponds to the Fourier transform of the displayed SLM phase \cite{Hecht2002}. For the trivial case of $l=0$, we recover a completely uniform phase pattern and thus a simple reflection of the incoming light, yielding a Gaussian mode [cf. Fig. \ref{fig:1}(e)].

\subsection{Photoluminescence experiments}
We utilize a laser with an energy of \SI{1.94}{\electronvolt} (\SI{639}{\nano\meter}) to generate the IXs in the \ce{MoSe2}/\ce{WSe2} heterostructures. The samples are measured inside a closed-cycle cryostat with a bath temperature of \SI{1.7}{K}. The excitation and detection energies are separated from each other by a \SI{650}{\nano\meter} long-pass dichroic mirror. Moreover, we utilize a long-pass filter with the same wavelength as well as a band-pass filter with a width of \SI{40}{\nano\meter} around \SI{880}{\nano\meter} (\SI{1.41}{\electronvolt}). For the first set of experiments [\crefs[(b)]{fig:1}, \sref[(a), (b)]{fig:2}, and \sref[(a)]{fig:3}], we utilize a CMOS camera to detect the corresponding PL at \SI{1.41}{\electronvolt}. For the hyperspectral characterizations [\crefs[(b-f)]{fig:3} and \sref[(d)]{fig:4}], we mount a movable multi-mode fiber with a core diameter of \SI{50}{\micro\meter} in the image plane, allowing for a spectrally resolved detection of the PL at different positions on the sample.

\subsection{Time-resolved measurements}
For determining the IX diffusion constants, we perform time-resolved photoluminescence measurements. To this end, the utilized excitation laser diode is switched to a pulsed mode by feeding a Transistor-Transistor Logic (TTL) pulse to its control input. We collect time-resolved photoluminescence signals by time tagging the TTL pulse for the excitation laser as well as the count pulses from an avalanche photodiode with the help of a Time-Correlated Single Photon Counting (TCSPC) module. This way, we acquire the relative timing of photon events with respect to the timing of the laser pulse [Supporting Fig. S7].

\section{Acknowledgments}
The authors gratefully acknowledge very productive discussions with A. Knorr and S. Meyer as well as the German Science Foundation (DFG) for financial support via Grants HO 3324/16-1, No. 290642686, 443274199, and 556436549 (WU 637 4-2, 7-1, 8-1), and the clusters of excellence MCQST (EXS-2111) and e-conversion (EXS-2089), and the priority program 2244 (2DMP) via HO3324/13-2. K.W. and T.T. acknowledge support from the JSPS KAKENHI (Grant Numbers 21H05233 and 23H02052), the CREST (JPMJCR24A5), JST and World Premier International Research Center Initiative (WPI), MEXT, Japan. 

\bibliography{bib}
\end{document}


\title{\texorpdfstring{Supporting Information\\[2ex]\MainTitle}{Supporting Information - \MainTitle}}
\def\wsi{Walter Schottky Institute and Physics Department, TU Munich, Am Coulombwall 4a, 85748 Garching, Germany}
\def\mcqst{Munich Center for Quantum Science and Technology (MCQST), Schellingstr. 4, 80799 Munich, Germany}
\def\berlin{Institute for Theoretical Physics, Nonlinear Optics and Quantum Electronics, Technical University of Berlin, 10623 Berlin, Germany}
\def\muenster{Institute of Physics, University of Münster, Wilhelm-Klemm-Str. 10, 48149 Münster, Germany}
\def\mana{Research Center for Materials Nanoarchitectonics, National Institute for Materials Science, 1-1 Namiki, Tsukuba 305-0044, Japan}
\def\nims{Research Center for Electronic and Optical Materials, National Institute for Materials Science, 1-1 Namiki, Tsukuba 305-0044, Japan}

\author{Mirco Troue}

\author{Johannes Figueiredo}
\author{Gabriel Mittermair}
\author{Jonas Kiemle}
\author{Sebastian Loy}\affiliation{\wsi}\affiliation{\mcqst}
\author{Hendrik Lambers}\affiliation{\muenster}


\author{Takashi Taniguchi}\affiliation{\mana}
\author{Kenji Watanabe}\affiliation{\nims}

\author{Ursula Wurstbauer}\affiliation{\muenster}
\author{Alexander W. Holleitner}\email[Electronic address: ]{holleitner@wsi.tum.de}\affiliation{\wsi}\affiliation{\mcqst}
\date{\today}

\maketitle

\section{S1 Creation of Laguerre-Gaussian excitation profiles}
We utilize a phase-only Liquid Crystal on Silicon (LCoS) spatial light modulator (SLM) to shape the excitation laser beam into a Laguerre-Gaussian (LG) mode. The size of the created mode changes with the topological order $l$ of the mode and follows a square root dependence.

\begin{figure}[h!]
    \centering
    \includegraphics{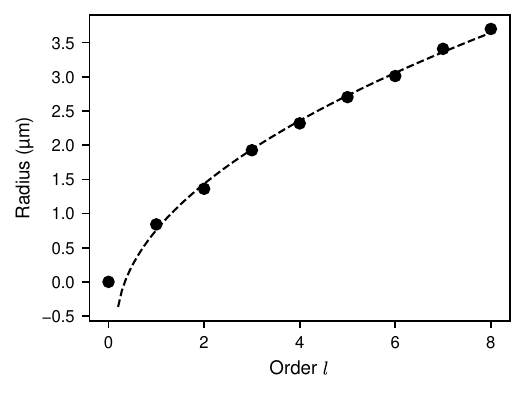}
    \caption{Radius of the reflected Laguerre-Gaussian mode for different topological orders $l$ as measured in the back reflection from the bare \ce{Si}/\ce{SiO2} substrate. The dashed line depicts the theoretically expected square root dependence.}
    \label{fig:S1}
\end{figure}

\clearpage
\section{S2 Spatial analysis of interlayer exciton photoluminescence from sample 2}
The emitted interlayer exciton photoluminescence is analyzed in a CMOS camera to resolve spatial features and the extent of the ensemble. A larger expansion as well as filling up of the center region with excitons is observed with increasing excitation powers. In addition to sample 1, another h-BN encapsulated vdW-heterostructure is investigated. Figure S2 shows the PL emission of sample 2 with an LG mode with order $l=4$ for varying powers. Again, we observe the same characteristics as for sample 1. Higher excitation powers lead to a relative increase of the exciton densities in the center as well as increased expansion of the ensemble. Sample 2 shows a stronger exciton trapping effect in the center. However, a more uneven distribution of PL emission is observed, which we attribute to the local fluctuations in the underlying potential landscape. These potential fluctuations, including local strain changes as well as altered interlayer coupling, arise, for example, from impurities or defects in the investigated heterostructures [9].

\begin{figure}[h!]
    \centering
    \includegraphics{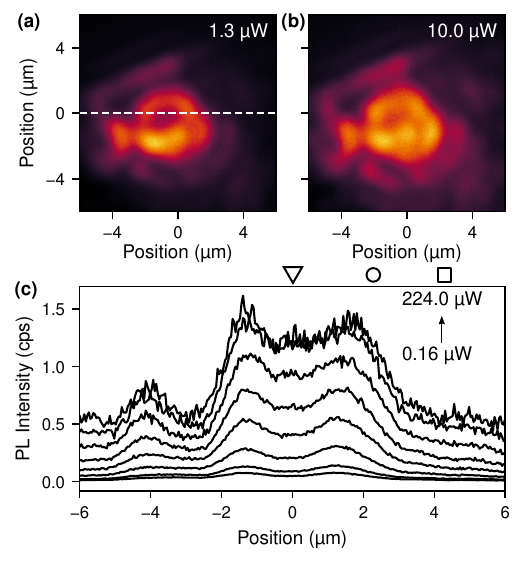}
    \caption{Interlayer exciton photoluminescence images in the lateral sample plane for varying powers on sample 2. (a) and (b): Exemplary camera images of the photoluminescence for two different powers. (c) Intensity profiles along the dashed line shown in (a).}
    \label{fig:S2}
\end{figure}

\clearpage
\section{S3 Hyperspectral analysis of interlayer exciton photoluminescence}
We collect the emitted light in the detection path from the image plane with the help of an optical multimode fiber. The spatial resolution with the multimode fiber is sufficient to clearly distinguish the emission from inside or outside the excitation profile. The subsequent analysis with a spectrograph yields hyperspectral data of the emission.

\begin{figure}[h!]
    \centering
    \includegraphics{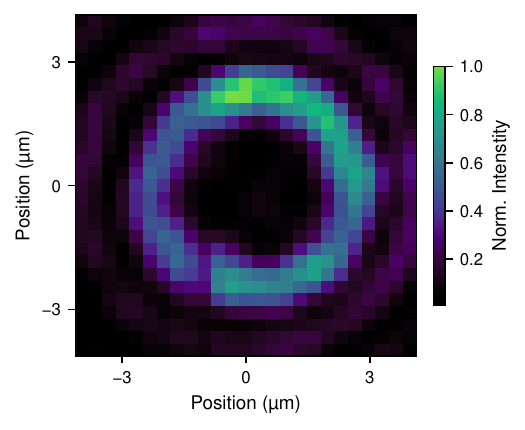}
    \caption{Spatial scan of the reflected laser illumination from the bare substrate in the detection plane utilizing a multimode fiber with a \SI{50}{\micro\meter} core diameter for collection of the emission and an avalanche photodiode for detection.}
    \label{fig:S3}
\end{figure}

\begin{figure}[h!]
     \centering
     \includegraphics[width=\columnwidth]{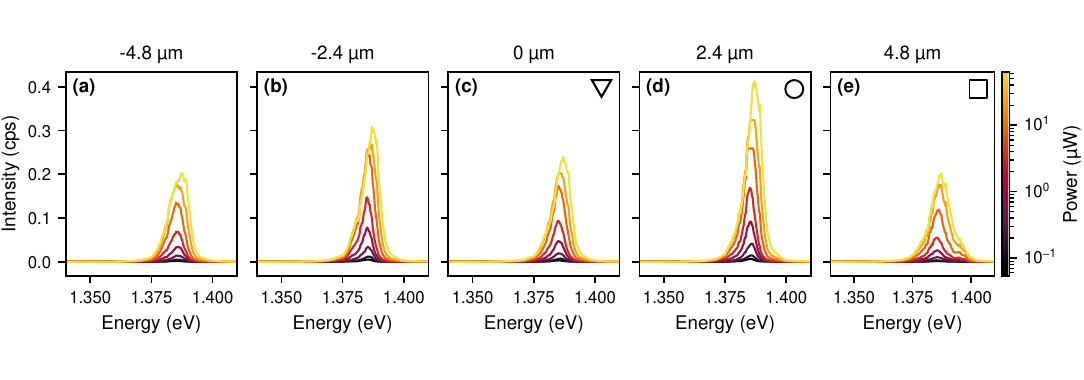}
     \caption{Extracted single spectra from hyperspectral measurements as shown in Fig. 3. (a) - (e): Power-dependent spectra for different positions across the exciton trap.}
     \label{fig:S4}
\end{figure}

\begin{figure}[h!]
  \centering
  \includegraphics{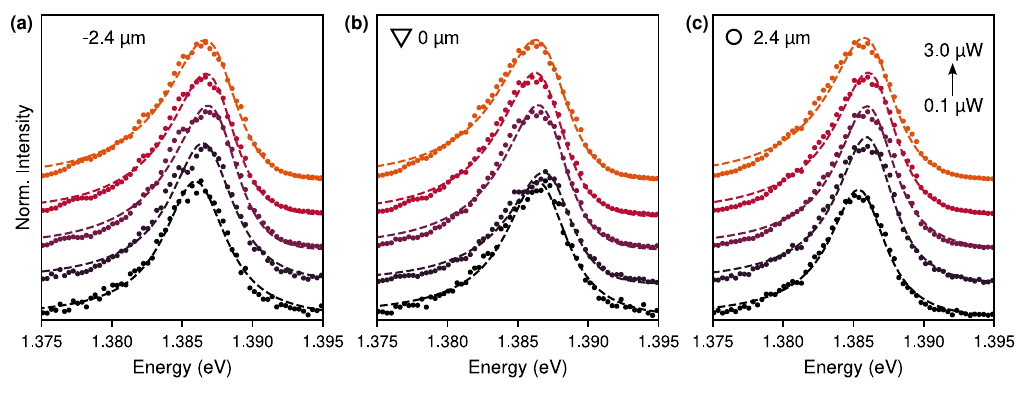}
  \caption{Single spectra and skewed Lorentzian fits from hyperspectral measurements as shown in Fig. 3 for the five lowest investigated powers. (a) - (c): Power-dependent spectra for different positions across the exciton trap.}
  \label{fig:S5}
\end{figure}

\begin{figure}[h!]
 \centering
 \includegraphics[width=\columnwidth]{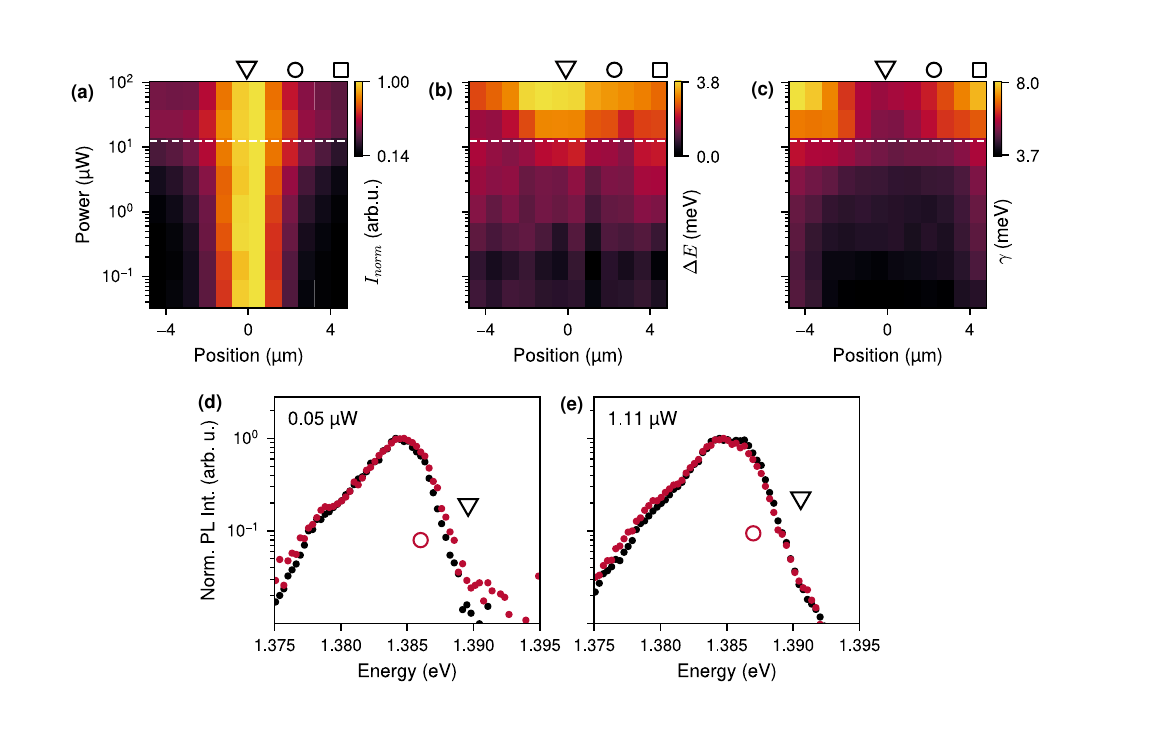}
 \caption{Hyperspectral properties of IX photoluminescence under Gaussian excitation. (a) - (c): Normalized intensity $I_{\mathrm{norm.}}$, energy shift $\Delta E$, and linewidth $\gamma$ of the exciton emission for varying excitation power and position. The white dashed lines depict the power corresponding to the highest investigated excitation power density for the $l=4$ excitation. For that excitation scheme, the laser intensity is distributed across the whole perimeter of the ring, such that the effective power density is lower than for the case of a Gaussian excitation. (d) and (e): Logarithmic spectra for exemplary excitation powers at the center position (black) and \SI{2.4}{\micro\meter} shifted to the right (red).}
 \label{fig:S6}
\end{figure}

\clearpage
\section{S4 Radiative interlayer exciton lifetime}
Time-resolved luminescence measurements use a Time-Correlated Single Photon Counting (TCSPC) module in combination with a pulsed laser excitation. A distinction between center position and position on the excitation ring is possible due to the aforementioned collection with a movable multimode fiber. Comparing the photoluminescence lifetime measurement in the center with a measurement on the ring shows only a small variation in the lifetime. A decay time of \SI{19.29(0.87)}{\nano\second} is extracted for the lowest power shown when measuring on the ring. We are able to determine the diffusion constant from the mean squared displacement $\langle r^2 \rangle \approx \SI{0.46}{\micro\meter\squared}$ obtained from the 1D cuts shown in Fig. 2.
\begin{equation}
    D = \dfrac{\langle r^2 \rangle}{2\tau} \approx \SI{0.12}{\centi\meter\squared\per\second}
\end{equation}

\begin{figure}[h!]
    \centering
    \includegraphics[width=\columnwidth]{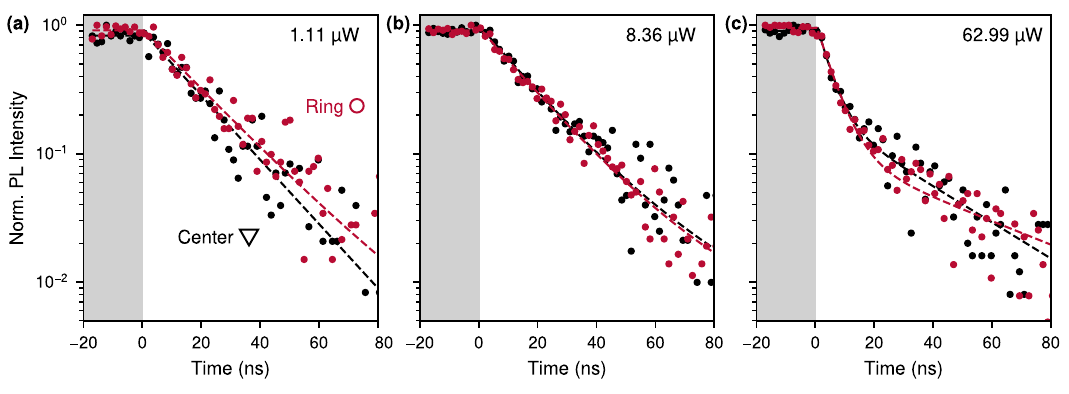}
    \caption{Spatially resolved radiative IX lifetime for varying excitation powers. (a) - (c) Decay histograms for \SI{1.11}{\micro\watt}, \SI{8.36}{\micro\watt}, and \SI{62.99}{\micro\watt} excitation power at the center position $\triangledown$ (black) and on the excitation ring $\circ$ (red) collected with an avalanche photodiode. The dashed lines showcase mono-exponential fits. For (b) and (c), the decay is modeled with an additional faster decay channel. The laser is turned off at \SI{0}{\nano\second}.}
    \label{fig:S7}
\end{figure}